\documentclass[aps,prd,twocolumn,preprintnumbers,superscriptaddress,10pt]{revtex4-1}
\usepackage{epsf,color,amsmath,amssymb,amsfonts}
\usepackage{enumerate}
\usepackage{hyperref}
\usepackage{graphicx}
\usepackage{psfrag}
\usepackage{dcolumn,bm}

\newcommand{\bea}{\begin{eqnarray}}
\newcommand{\beal}[1]{\begin{eqnarray}\label{#1}}
\newcommand{\eea}{\end{eqnarray}}
\newcommand{\be}{\begin{equation}}
\newcommand{\bel}[1]{\begin{equation}\label{#1}}
\newcommand{\ee}{\end{equation}}

\newcommand{\nn}{\nonumber}
\newcommand{\bit}{\begin{itemize}}
\newcommand{\eit}{\end{itemize}}
\newcommand{\ben}{\begin{enumerate}}
\newcommand{\een}{\end{enumerate}}

\def\t{\tilde}

\begin{document}
\onecolumngrid

\title{Testing the membrane paradigm with holography}

\author{Jan de Boer}
%\email{j.deboer@uva.nl}
\affiliation{Instituut voor Theoretische Fysica, Universiteit van Amsterdam \\
Science Park 904, 1090 GL Amsterdam, The Netherlands}

\author{Michal P. Heller}
%\email{mheller@perimeterinstitute.ca}
%\altaffiliation[On leave from: ]{\emph{National Centre for Nuclear Research,  Ho{\.z}a 69, 00-681 Warsaw, Poland.}}
\affiliation{Perimeter Institute for Theoretical Physics, Waterloo, Ontario N2L 2Y5, Canada}
\affiliation{National Centre for Nuclear Research,  Ho{\.z}a 69, 00-681 Warsaw, Poland}
\affiliation{Instituut voor Theoretische Fysica, Universiteit van Amsterdam \\
Science Park 904, 1090 GL Amsterdam, The Netherlands}

\author{Natalia Pinzani-Fokeeva}
%\email{N.PinzaniFokeeva@uva.nl}
\affiliation{Instituut voor Theoretische Fysica, Universiteit van Amsterdam \\
Science Park 904, 1090 GL Amsterdam, The Netherlands}

\begin{abstract}
One version of the membrane paradigm states that as far as outside observers are concerned, black holes can be replaced by a dissipative membrane with simple physical properties located at the stretched horizon. We demonstrate that such a membrane paradigm is incomplete in several aspects. We argue that it generically fails to capture the massive quasinormal modes, unless we replace the stretched horizon by the exact event horizon, and illustrate this with a scalar field in a BTZ black hole background. We also consider as a concrete example linearized metric perturbations of a five-dimensional AdS-Schwarzschild black brane and show that a spurious excitation appears in the long-wavelength response that is only removed from the spectrum when the membrane paradigm is replaced by ingoing boundary conditions at the event horizon. We interpret this excitation in terms of an additional Goldstone boson that appears due to symmetry breaking by the classical solution ending on the stretched horizon rather than the event horizon.
\end{abstract}

\maketitle

\section{Introduction}
Describing black holes in general relativity is a complicated endeavor and, as a consequence, various approximation schemes have been developed over the years. One of them is the membrane  paradigm~\cite{Damour:1978cg}, in which the black hole is replaced by a 
%It is well known that astrophysical black holes are quite complicated and intractable objects in the realm of the full  general relativity description. In order to overcome such difficulties  one necessarily has to rely on  
%approximation schemes of increasing accuracy which have been extensively developed in the past decades. Among those, 
%the membrane paradigm \cite{Damour:1978cg}
 % replaces the black hole by a 
simple dissipative membrane situated at the stretched horizon, i.e. a very small distance 
from the event horizon \footnote{Much smaller than any other scale in the problem.}. In the astrophysical context, this membrane approximation %is endowed with particularly simple physical properties, such as constant AC conductivity and 
has been applied to %This simple prescription of ignoring the near horizon details of  astrophysical problems has been applied e.g. for 
the study of the magnetosphere of a black hole surrounded by a magnetized accretion disk \cite{Macdonald} and jets \cite{Penna:2013rga}, see \cite{Thorne:1986iy} for many more applications and further references. More recent interest in the membrane paradigm has been stimulated by the discovery of holography \cite{Maldacena:1997re,Gubser:1998bc,Witten:1998qj} and what eventually became known as the fluid-gravity duality \cite{Bhattacharyya:2008jc,Hubeny:2011hd}, see \cite{Damour:2008ji} for a review of the membrane paradigm in this context.

It has, in our opinion, always been somewhat mysterious to what extent, and in what sense this membrane paradigm is an accurate statement. Does it fail to reproduce some aspects of the correlation
functions measured at infinity? Are there any undesired effects stemming from the fact that the membrane lives on the local stretched horizon (timelike hypersurface), rather than on the null event horizon, a teleologically defined object?
%Does the membrane really live at the local stretched horizon which is a timelike hypersurface or should it be thought of as living on the event horizon, which is a global (teleologically defined) notion?
%A well known application of the AdS/CFT correspondence \cite{Maldacena:1997re,Gubser:1998bc,Witten:1998qj} is to use it to compute finite temperature transport coefficients and analyze more general non-equilibrium phenomena in strongly coupled field theories. Dissipation arises
%in gravity by imposing ingoing boundary conditions at the black hole horizon, which allows the black hole to
%absorb energy. This is to be contrasted with the membrane paradigm \cite{Damour:1978cg,Thorne:1986iy,Parikh:1997ma}, which we take to be the statement that as far as outside observers
%are concerned, one can replace the black hole by a dissipative membrane situated at the stretched horizon, i.e. a very small distance\footnote{i.e. much smaller than any other scale in the problem.}
%away from the event horizon.
%This membrane is endowed with particularly simple physical properties, such as constant AC conductivity.
In this letter we show that the answers to both questions are affirmative. For concreteness, we mostly work in the context of the AdS/CFT 
correspondence \cite{Maldacena:1997re,Gubser:1998bc,Witten:1998qj}, but our results apply much more generally. 

Following \cite{Parikh:1997ma,Iqbal:2008by}, the membrane paradigm will be for us a relation between the
radial derivative of the field and the field itself which 
would naively encode the ingoing boundary condition at the event horizon if it were evaluated there. 
To derive it, we 
consider a probe scalar field in the background of a (d+1)-dimensional AdS-Schwarzschild black brane
%\be
%\label{eq.AdS-Schw}
%ds^2 = \frac{L^2du^2}{4 u^2 f(u)} - \frac{(4\pi T L/d)^{2}}{u} f(u) dt^{2} + \frac{(4\pi T L/d)^{2}}{u} d\vec{x}^{2},
%\ee
\be
\label{eq.AdS-Schw}
ds^2 = \frac{du^2}{4 u^2 f(u)} - \frac{(4\pi T/d)^{2}}{u} f(u) dt^{2} + \frac{(4\pi T/d)^{2}}{u} d\vec{x}^{2},
\ee
where $f(u) = 1 - u^{d/2}$, the horizon is at $u = 1$ and we set the AdS radius to unity. A scalar field $\phi$ has two possible solutions near the horizon
\bea
\label{eq.phinh}
\phi = && e^{- i \omega t + i \vec{k} \cdot \vec{x}} \, \Bigg\{ c_{out} (1-u)^{i \tilde{\omega}/2} \Big(1 + \alpha_{1} (1- u) + \ldots\Big)  + \nonumber\\
&& c_{in} (1 - u)^{-i \tilde{\omega} /2} \Big(1 + \beta_{1} (1-u) + \ldots\Big) \Bigg\},
\eea
where, to keep the notation simple, we defined
$\tilde{\omega} = \omega / (2\pi T)$.
Let us emphasize that the form of the expansion~(\ref{eq.phinh}) is valid near the horizon of {\it any} non-extremal black hole.
The universal leading terms $(1-u)^{\pm i \tilde{\omega}/2}$ follow from the fact that the 
near-horizon region of any finite-temperature black hole is Rindler spacetime. The values of the coefficients $\alpha_{i}$ and $\beta_{j}$ of the subleading terms are non-universal and depend on
the number of dimensions, the mass of the field and its momentum.

%The universal form of the leading order terms in eq. \eqref{eq.phinh}, i.e. $(1-u)^{\pm i \tilde{\omega}/2}$, follows from the fact that the near-horizon region of any finite-temperature black hole is Rindler spacetime. The dependence on 
%the number of dimensions, mass of the field and its momentum is hidden in the form of the coefficients $\alpha_{i}$ and $\beta_{j}$ that multiply the subleading terms.

%When calculating the retarded Green's function of an operator dual to $\phi$, or the quasinormal modes which appear as poles of this Green's function, one imposes the ingoing boundary condition $c_{out} = 0$ and supplements it with suitable boundary conditions at infinity. For the advanced Green's function the relevant boundary condition is the purely outgoing one. 

One typically is interested in imposing ingoing boundary conditions at the event horizon, i.e. $c_{out} = 0$ (or
$c_{in} = 0$ for purely outgoing boundary conditions). A suggestive but not entirely accurate way of 
rewriting these boundary conditions is
\be
\label{eq.mpscalar}
2 (1-u)\frac{\partial_{u} \phi}{\phi} \Bigg|_{u = 1} = i \, \tilde{\omega} \, \sigma,
\ee
where $\sigma = 1$ corresponds to purely ingoing and $\sigma = -1$ to purely outgoing modes. Eq. \eqref{eq.mpscalar}, as a ratio of the momentum
of the field to the field itself, can be reinterpreted as a universal response function characterizing non-extremal horizons and yet having a particularly simple form. By analogy with the electromagnetic case, we will be referring to $\sigma$ as the horizon conductivity or the membrane coupling. Note though that, strictly speaking, for generic solutions of the field equations the expression \eqref{eq.mpscalar} is ill-defined at the horizon. 

The statement of the membrane paradigm that we are going to adopt follows \cite{Parikh:1997ma,Iqbal:2008by} and earlier work and amounts to keeping $\sigma$ fixed and equal to $1$ and viewing eq. \eqref{eq.mpscalar} not as the
response of the event horizon, a null surface residing at $u = 1$, but rather of a timelike membrane located at $u = u_{\delta} = 1 - \delta$ with $\delta$ very small. One might have thought that for sufficiently small $\delta$ the membrane paradigm always \emph{effectively} imposes ingoing boundary conditions on the event horizon. Surprisingly, this turns out not to be the case.

To demonstrate this, we follow \cite{Faulkner:2010jy,Heemskerk:2010hk,Nickel:2010pr}
and decompose the bulk spacetime into a near-horizon region (``IR region'' with $1 \geq u > u_{\delta}$) and the rest (``UV
region'' with $0 \leq u < u_{\delta}$). In this approach, eq. \eqref{eq.mpscalar} is supposed to model
 the dynamics of the ``IR region'' and the role of the fields living in the ``UV
region'' is to propagate the membrane paradigm response to infinity ($u = 0$), where the correlation functions of the dual field theory
are defined. It will be convenient to impose eq. \eqref{eq.mpscalar} in two steps: we first solve in the Fourier space for the fields in the UV-region
subject to Dirichlet boundary conditions for the fields at two different slices with fixed $u$, 
in our case at $u = 0$ and at $u = u_{\delta}$. Next, we remove one of the Dirichlet boundary conditions by fixing the value of the field at $u=u_{\delta}$ requiring 
eq. \eqref{eq.mpscalar}. %In general, the solutions to the first ``double Dirichlet'' problem are not unique, as they can contain an arbitary dependence on massive Kaluza-Klein modes, but these modes are not relevant for the discussion in this paper and we will ignore them. 
Additional complications are going to arise when the bulk fields transform nontrivially under some local symmetry. The solutions in the ``UV region'' will then also depend on certain gapless degrees of freedom, which arise as the Goldstone bosons of the symmetries broken by the classical solution and ensure that dual operators are conserved currents \cite{Nickel:2010pr}.

\section{Massive quasinormal modes}

We start by investigating in detail the scalar field case discussed in the introduction. In holography, solutions of the equations of motion for a scalar satisfying ingoing boundary conditions at the event horizon encode the retarded two-point function of a dual scalar operator in the thermal state. Conversely, imposing the outgoing boundary conditions leads to the advanced two-point function. We take the frequency $\tilde{\omega}$ to be a complex number in order to accommodate for the quasinormal modes, which are the poles of the retarded two-point function.

In our setup we want to replace the ingoing boundary conditions at the event horizon by eq.~\eqref{eq.mpscalar} imposed at the stretched horizon $u=u_{\delta}$. For the moment, we ignore the form of $\phi$ at infinity and focus on its near-horizon behavior, which is given in full generality by eq. \eqref{eq.phinh}. If we apply the membrane paradigm relation to eq. \eqref{eq.phinh}, we readily 
obtain that
\bea
\label{eq.cout2cin}
&& c_{out} / c_{in} = (1-u_\delta)^{-i \tilde{\omega}} \times \nonumber \\
&& \frac{(1-\sigma) \tilde{\omega} + \beta_{1} \left(2 i + (1-\sigma ) \tilde{\omega} \right) (1-u_\delta) + \ldots}{(1+\sigma) \t\omega + \alpha_{1} \left( -2 i + (1 + \sigma) \tilde{\omega} \right) (1-u_\delta) + \ldots } \,.
\eea
Using $\sigma=1$ and keeping only the leading order terms, eq. \eqref{eq.cout2cin} reduces to
\be
\label{eq.cout2cinnh}
c_{out} / c_{in} = (1-u_\delta)^{1-i\tilde{\omega}} \times \frac{i \beta_{1}}{\tilde{\omega}}.
\ee
It is easy to see that for values of $\tilde{\omega}$ such that $\Im(\tilde{\omega}) > -1$, this formula 
has the desirable effect, i.e. it leads to $|c_{out} / c_{in}| \ll 1$ for $u_{\delta} \rightarrow 1$, but for 
$\Im(\tilde{\omega}) < -1$ it does not, it effectively leads to outgoing boundary conditions $|c_{out} / c_{in}| \gg 1$ as $u_{\delta} \rightarrow 1$ instead. Note that this holds no matter how close to the event horizon the membrane is. This implies that using the membrane paradigm on the stretched horizon with $\sigma=1$ only yields a good approximation to the retarded Green's function if $\Im(\tilde{\omega}) > -1$, whereas for $\Im(\tilde{\omega}) < -1$ we obtain instead an approximation to the advanced Green's function. This, in particular, means that, at best, the membrane paradigm will reveal only a few of the lowest lying quasinormal modes if any.

The root of this discrepancy is that $\sigma = \pm 1$ is formally correct only at the event horizon and away from it is slightly different. This ``flow'' of the membrane conductivity can be constructed perturbatively in the $1-u_{\delta}$ expansion for $u_{\delta}$ sufficiently close to 1. The form of eq. \eqref{eq.cout2cin} implies that we may be able to cover somewhat wider range of imaginary parts of the frequency if we 
include a finite number of these corrections in $\sigma$. It is clear though that in order to cover
the whole complex frequency plane the value of the \emph{exact} membrane conductivity at any $u_{\delta} \neq 1$ is required, 
which is equivalent to knowing the whole solution with the ingoing boundary conditions at the event horizon. However, this defies the purpose of introducing the membrane paradigm.

We would like to emphasize that the above analysis relies only on the universal near-horizon region and hence is 
widely applicable, also to gravitational perturbations and asymptotically flat black holes.

\begin{figure*}
\begin{tabular}{cc}
\includegraphics[width=0.42 \textwidth]{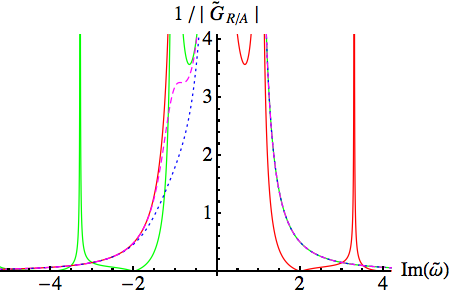}
\quad & \quad
\includegraphics[width=0.42 \textwidth]{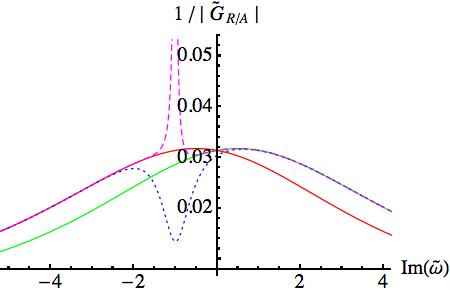}
\end{tabular}
\vspace{-0.1cm}
\caption{Absolute value of the inverse of the retarded/advanced Green's function (green/red) and the membrane paradigm approximations of the former at $u_{\delta} = 0.9$ (blue) and $u_{\delta} = 0.999$ (magenta) as a function of $\Im{(\tilde{\omega})}$ for $\Re{(\tilde{\omega})} = 0$ (left) and $\Re{(\tilde{\omega})} = 5$ (right). One can clearly see that the stretched horizon approximation works for $\Im{(\tilde{\omega})} > -1$, whereas for $\Im{(\tilde{\omega})} < -1$ it leads to the advanced Green's function, in line with the approximation in eq. \eqref{eq.cout2cinnh}. Zeros of the green curve correspond to the locations of the quasinormal modes, as given by eq. \eqref{eq.qnmpos}, and lie beyond the range of applicability of the membrane paradigm.
}
\label{fig1}
\end{figure*}

Let us now demonstrate our findings for the exactly soluble case of a massless scalar field in $2+1$ dimensions \footnote{We repeated the same calculation in (4+1)-dimensions and found similar behavior.} (the BTZ black brane background) in Einstein gravity with a negative cosmological constant \cite{Banados:1992wn,Son:2002sd}. For simplicity, we will set the momentum $k$ to 0 and write the bulk field as $\phi(u) e^{- i \omega t}$. The scalar is dual to an operator $\cal O$ of conformal dimension $\Delta = 2$ in the dual (1+1)-dimensional conformal field theory. Its retarded/advanced Green's function is given by
\be
\label{eq.GRA}
G_{R/A} \Big/ \left(\frac{\pi T^{2}}{4 G_{N}}\right) = \tilde{G}_{R/A}= \frac{\phi_{1}}{\phi_{0}} - \frac{1}{2} \tilde{\omega}^{2},
\ee
where $\phi_{0,1}$  denote the (non)normalizable behavior of $\phi$ near the AdS boundary
\be
\phi(u) = \phi_{0} + (\phi_{1} - \frac{1}{4} \tilde{\omega}^{2} \phi_{0} \log{u}) u + \ldots
\ee
and the solutions obey ingoing/outgoing boundary conditions at the event horizon. The quasinormal modes appear as the poles of $G_{R}$ and have frequencies
%which is proportional to
%\be
%\tilde{G}_{R} = \frac{i \, \tilde{\omega}}{2} - \frac{\tilde{\omega}^{2}}{4} - \frac{\gamma \, \tilde{\omega}^{2}}{2} - \frac{\tilde{\omega}^{2}}{2} \, \psi_{0}\left(1-\frac{i \, \tilde{\omega} }{2}\right)
%% - i \, \tilde{\omega} + \gamma \, \tilde{\omega}^{2} + \tilde{\omega}^2 \, \psi_{0}\left(1-\frac{i \, \tilde{\omega} }{2}\right),
%\ee
%with $\gamma$ the Euler constant and $\psi_{0}(x)=\Gamma'(x)/\Gamma(x)$, and diverges (with single pole singularities) at
\be
\label{eq.qnmpos}
\tilde{\omega} = - 2 \, i \, n \quad \mathrm{for} \quad n = 1, 2, \ldots
\ee
In order to test the membrane paradigm, we solve the scalar field equation of motion for the configuration obeying Dirichlet boundary conditions \footnote{For a massive field, we would typically demand that at $u = 0$ the leading fall-off of the solution is fixed.} at $u=0$ and $u=u_{\delta}$,
\be
\label{eq.ddscalar}
\phi(u = 0) = \phi_{0} \quad \mathrm{and} \quad \phi(u = u_{\delta}) = \phi_{\delta},
\ee
and subsequently use eq. \eqref{eq.mpscalar} evaluated at $u_{\delta}$ to express $\phi_{\delta}$ in terms of $\phi_{0}$. This, in turn, determines $\phi_1$, which is enough to evaluate eq. \eqref{eq.GRA}. The results are summarized in Fig.~\ref{fig1} and nicely confirm
the general result obtained above, as we indeed see that the retarded Green's function is not well-approximated for 
$\Im{(\tilde{\omega})} < -1$ and the advanced Green's function is not well-approximated for $\Im{(\tilde{\omega})} > 1$. This, in particular, implies that none of the quasinormal modes in this setup are captured by the membrane paradigm.
\section{Sound waves}
Let us now reconsider long-wavelength gravitational perturbations of a (4+1)-dimensional AdS black brane in Einstein gravity from the point of view of the membrane paradigm. Earlier approaches to this problem include \cite{Kovtun:2003wp,Starinets:2008fb,Iqbal:2008by,Nickel:2010pr,Faulkner:2010jy,Emparan:2013ila,Pinzani-Fokeeva:2014cka}. For definiteness, we will focus on the sound channel perturbations with momentum in the
$x$-direction for which the non-vanishing metric variations are
\be
\label{eq.gravpert}
\delta  h_{tt},\,\,\, \delta  h_{tx},\,\,\, \delta h_{xx} \,\,\, \mathrm{and} \,\,\, \delta  h_{aa}=\frac{1}{2}(\delta  h_{yy}+\delta h_{zz}).
\ee
In fact, demanding that $\delta  h_{\mu u}  = 0$ is just a convenient gauge choice. It will also be useful to 
redefine (\ref{eq.gravpert}) in the following way
\be \label{eq.redef}
H_{\mu\nu}(u) e^{- i \omega t + i k x}:=|g^{\mu\rho}|\delta h_{\rho\nu},
\ee
where $g_{\mu \nu}$ is the black brane metric \eqref{eq.AdS-Schw}.

The standard approach \cite{Kovtun:2005ev} in dealing with the gravitational perturbations (\ref{eq.gravpert}) is to introduce the gauge-invariant variable 
\bea
Z(u)&=&2k^2f(u)H_{tt}(u)+4 \omega kH_{tx}(u)+2\omega^2H_{xx}(u)+\nn\\
&&+H_{aa}(u)\left(k^2(1+u^2)-\omega^2\right).
\eea
Using all linearized Einstein equations one obtains a decoupled second order ordinary differential equation for $Z$, and therefore the problem of finding the retarded stress tensor correlator in the sound channel is completely analogous to the scalar field case studied in the previous section. Hence, to test the membrane paradigm we once more impose the universal relation \eqref{eq.mpscalar} on a stretched horizon $u_{\delta}$ with $\phi$ simply replaced by $Z$.
 
A generic solution for $Z$ can be found analytically order by order in a hydrodynamic expansion
\bea\label{sol}
Z(u)&&=c_{in}(1-u^2)^{-i\t \omega/2}\left(X_0(u)+\lambda\, X_1(u)+\dots\right)+\nn\\
&&+c_{out}(1-u^2)^{i\t\omega/2}\left(Y_0(u)+\lambda\, Y_1(u)+\dots\right),
\eea 
where $\lambda$ is a bookkeeping parameter counting powers of $\t\omega \ll 1$ and $\t k = k/(2\pi T)\ll1$. Requiring $c_{out}=0$ selects the ingoing mode at the horizon. To leading order in $\lambda$ the solution reads
\be
X_{0}(u)=Y_{0}(u)=\frac{k^2(1+u^2)\!-\!3\omega^2}{5k^2\!-\!3\omega^2},
\ee
which, together with the membrane paradigm \eqref{eq.mpscalar}, give a relation  
between the outgoing and ingoing coefficients $c_{out}$ and $c_{in}$ on the stretched horizon. By direct computation in a near horizon expansion 
the analogue of eq.~\eqref{eq.cout2cin} now takes the form
\bea\label{ratio}
&&c_{out}/c_{in}= (1-u_{\delta})^{-i \t\omega }\frac{(1-\sigma) }{2^{i\t\omega}(\sigma +1)}+\nn\\
&&+ (1-u_{\delta})^{1-i\t\omega} \frac{ g({\t\omega},{\t k},\sigma) }{ (\sigma +1)^2 \t\omega\left(3\t\omega ^2-2 \t k^2  \right)}+\dots,
\eea
where $g$ is an analytic function of $\t \omega$, $\t k$ and the membrane coupling $\sigma$. One can clearly see that for small $\t \omega$ and $\t k$ one indeed obtains from \eqref{ratio} a very small ratio of $c_{out}/c_{in}$ unless $\tilde{\omega} = \pm \sqrt{2/3} \, \tilde{k}$. With some work, one can also determine the approximate retarded Green's function, and from its poles
one obtains two branches of solutions, which for small enough $\delta$ read
\be
\label{eq.omegasw}
\tilde{\omega} = \pm \sqrt{\frac{1}{3}} \, \tilde{k} + \mathcal{O}(\tilde{k}^{2})
\ee
and
\be
\label{eq.omegaspurious}
\tilde{\omega} = \pm \sqrt{\frac{2}{3}} \, \tilde{k} + \mathcal{O}(\tilde{k}^{2}).
\ee
The mode \eqref{eq.omegasw} is just the standard hydrodynamic sound wave, whereas the second one is spurious, as it does not solve the linearized equations of hydrodynamics of the underlying microscopic theory and seizes to exist when one imposes the ingoing boundary condition at the event horizon. Moreover, the presence of a pole in the second term in eq.  \eqref{ratio} implies that the solution with the membrane paradigm boundary condition on a stretched horizon and with ingoing boundary conditions on the event horizon are not smoothly connected to each other for $\t\omega =\pm \sqrt{2/3}\,\t k$. Hence, the mode \eqref{eq.omegaspurious} has to be discarded. This yields one more  model-dependent restriction on the allowed frequencies.

In calculating the above leading order dispersion relations we kept $\sigma$ arbitrary and the result did not depend on the value of $\sigma$. This suggests that we can interpret both modes as arising from the ``UV region'' of the spacetime rather than being an intrinsic property of the membrane. Indeed, the emergence of hydrodynamical modes in the holographic context can be thought of as 
arising due to spontaneous symmetry breaking by the classical solution ending on the second boundary, i.e. the stretched or event horizon \cite{Nickel:2010pr}. 
To see this in our context, consider a double Dirichlet problem in the most general metric ansatz compatible with the symmetries of the
sound mode, i.e. besides the modes in \eqref{eq.gravpert} we also allow non-zero $H_{u t}$, $H_{u x}$ and $H_{u u}$.
These modes are defined as in (\ref{eq.redef}) except that it will turn out to be convenient to define
$\partial_u H_{uu}=u\sqrt{f(u)} \delta h_{uu}$.

It is easy to see that it is impossible to completely gauge away $H_{u t}$, $H_{u x}$ and $H_{u u}$ without modifying
the metric perturbations $H_{\mu\nu}$ at the two boundaries $u=0$ and $u=u_{\delta}$. In fact, one can gauge away 
$H_{u t}$, $H_{u x}$ and $H_{u u}$ up to the non-local ``Wilson-line'' like variables
\begin{subequations}
\bea
\label{eq.psit}
&&\psi_t({\t \omega},{\t k})=\int_0^{u_{\delta}}H_{tu}(u)du+i\,{\t \omega} \int_0^{u_{\delta}}\frac{H_{uu}(u)}{2f(u)^{3/2}}du,\quad\quad\\
\label{eq.psix}
&&\psi_{x}({\t \omega},{\t k})=\int_0^{u_\delta}H_{xu}(u)du-i\,{\t k} \int_0^{u_{\delta}}\frac{H_{uu}(u)}{2\sqrt{f(u)}}du, \quad \quad
\eea
\label{eq.goldstones}%
\end{subequations}
which we interpret, in line with \cite{ArkaniHamed:2002sp,Nickel:2010pr}, as the Goldstone bosons associated 
to the global symmetries broken by the classical solution \cite{toappear}. In addition, the constraint equations of general relativity are the 
equations of motion for the Goldstone bosons given by eq.~\eqref{eq.goldstones} and for $H_{u u}$.

The membrane paradigm couples the radial derivative of the metric to the metric itself with coupling strength~$\sigma$.
However, to the order we have been working at, the dispersion relations \eqref{eq.omegasw} and \eqref{eq.omegaspurious} 
do not depend on this coupling, suggesting that these modes arise from the Goldstone modes. To examine whether this is
indeed the case, we take arbitrary $H_{u t}$, $H_{u x}$ and $H_{u u}$ with all other metric perturbations equal to 
zero, and then make a gauge transformation to the radial gauge $H_{u t}=H_{u x}=H_{u u}=0$. This will turn on a non-trivial
metric perturbation on the stretched horizon given by
\begin{subequations}
\begin{eqnarray}
\label{eq.goldHtt}
H_{tt}(u_{\delta})&=&2\,i\,\t\omega\,\psi_t-\frac{(1+u_{\delta}^2)}{2f(u_{\delta})}H_{uu}(u_{\delta}),\\
 \label{eq.goldHtx}
H_{tx}(u_{\delta})&=&i\,\t\omega\,\psi_x-i\,\t kf(u_{\delta})\psi_t,\\
\label{eq.goldHxx}
H_{xx}(u_{\delta})&=&-2\,i\,\t k\,\psi_x+\frac{1}{2}\sqrt{f(u_{\delta})}H_{uu}(u_{\delta}),\\
\label{eq.goldHaa}
H_{aa}(u_{\delta})&=&\frac{1}{2}\sqrt{f(u_{\delta})}H_{uu}(u_{\delta}).
\end{eqnarray}
\label{eq.IRbdrycond}%
\end{subequations}
Strictly speaking, the solutions of the Einstein equation in the radial gauge with these boundary conditions and vanishing at infinity make sense only in the leading order of the derivative expansion. As the $\mu\nu$-components of the Einstein equations completely determine the bulk metric in the radial gauge, the remaining $u \nu$- and $uu$-components yield then non-trivial equations for the Goldstone modes $\psi_x$ and $\psi_t$, and
for $H_{uu}$. In the limit where the stretched horizon is very close to the event horizon, at leading order in $\tilde{k}$ and $\tilde{\omega}$ and after solving for $H_{uu}(u_{\delta})$, we arrive at the following two equations
\begin{subequations}
\bea
\label{const1}
&&\left(3 \, \tilde{\omega}^2-\tilde{k}^2 \right) \psi_x-\frac{3}{2} \, \tilde{\omega}\,\tilde{k}  \sqrt{f(u_{\delta})} \, \hat{\psi}_t=0,\quad\\
\label{const2}
&& \left(3 \, {\tilde \omega}^2 - 2 {\tilde k}^2 \, \right) \hat{\psi}_t + {\tilde \omega}\, {\tilde k}  \sqrt{f(u_{\delta})} \, \psi_x=0,
\eea
\label{constraint}%
\end{subequations}
where $\hat{\psi}_t = \sqrt{f(u_{\delta})} \psi_t$. This near-horizon redefinition would be natural if equations \eqref{const1} and \eqref{const2} would follow from an action principle in which the Goldstone bosons appear quadratically. The equations \eqref{const1} and \eqref{const2}, to leading order in $\delta$, directly lead to the sound waves \eqref{eq.omegasw} and the spurious mode \eqref{eq.omegaspurious}. Note that in the strict horizon limit the $\psi_{t}$ Goldstone decouples from the dynamics and one is only left with the hydrodynamic sound wave excitation. The decoupling of a Goldstone mode in this limit is related to the fact that the metric on the horizon is degenerate and hence causes a change in the symmetry breaking pattern.
\section{Conclusions}

In this letter we have studied the range of validity and applicability of the membrane paradigm, which we took to be a 
particular boundary condition eq. \eqref{eq.mpscalar} imposed at the stretched horizon, and which is supposed to represent the response
of the black hole to external perturbations. Though we mostly worked in the
context of the AdS/CFT correspondence, we expect our results to hold in more general gravitational setups as they rely on 
generic properties of horizons. Surprisingly,
we found that the membrane paradigm cannot be used to find the spectrum of quasinormal modes, and that it leads to a spurious
gapless sound-like mode which we interpreted in terms of Goldstone modes associated to
the symmetries broken by the classical solution ending on the stretched horizon. Both issues can in principle be resolved by an infinite fine-tuning of the membrane coupling or
conductivity $\sigma$, but this is equivalent to imposing ingoing boundary conditions directly at the horizon, and the main
purpose of the membrane paradigm was to not do that. 
Alternatively, in the case of sound modes, one can try to remove the problem by making
the cutoff $u_{\delta}$ $\tilde{k}$ and $\tilde{\omega}$ dependent, but superficially this once more requires significant fine-tuning.
%All of this suggests that it might be somewhat misguided to associate a simple
%dissipative sector to the sliver of spacetime between the stretched and the event horizon.

%\begin{acknowledgements}
%We would like to thank Alejandra Castro, Diego Hofman, Romuald Janik, Elias Kiritsis, Hong Liu, David Mateos and 
%Erik Verlinde for discussions and Roberto Emparan, Veronika Hubeny, Juan Jottar and Mukund Rangamani for discussions and their comments on the draft. MPH is supported
%by the Netherlands Organization for Scientic Research under the NWO Veni scheme (UvA) and by the National
%Science Centre under Grant No. 2012/07/B/ST2/03794 (NCNR). This work is part of the research programme of the
%Foundation for Fundamental Research on Matter (FOM), which is part of the Netherlands Organisation for Scientic Research (NWO).
%\end{acknowledgements}

\begin{acknowledgements}
We would like to thank A. Castro, R. Emparan, D. Hofman, V. Hubeny, R.~Janik, J. Jottar, E. Kiritsis, L. Lehner, H. Liu, D. Mateos, M. Rangamani, E.~Verlinde for discussions and correspondence. This work is supported
by the Netherlands Organization for Scientic Research under the NWO Veni scheme, by the National
Science Centre under Grant No. 2012/07/B/ST2/03794 and is part of the research programme of the
Foundation for Fundamental Research on Matter (FOM), which is part of the Netherlands Organisation for Scientic Research (NWO). Research at Perimeter Institute is supported by the Government of Canada through
Industry Canada and by the Province of Ontario through the Ministry of
Research \& Innovation.
\end{acknowledgements}

%%%%%%%%%%%%%%%

\bibliography{mempar_biblio}{}

\end{document}